\input harvmac
\def\lf{16\pi^2}

\def\frak#1#2{{\textstyle{{#1}\over{#2}}}}
\def\frakk#1#2{{{#1}\over{#2}}}
\def\pa{\partial}
\def\semi{;\hfil\break}
\def\ga{\gamma}

\def \th{\theta}
\def\Xtilde{\tilde X}
\def\sy{supersymmetry}
\def\sic{supersymmetric}

\def\mpla{{Mod.\ Phys.\ Lett.\ }{\bf A}}
\def\npb{{Nucl.\ Phys.\ }{\bf B}}
 
\def\prd{{Phys.\ Rev.\ }{\bf D}}

\def\plb{{Phys.\ Lett.\ }{\bf B}}

\def\Ph{\Phi}

\def\th{\theta}
\def\bM{M^*}

\def\bPh{\bar\Ph}
\def\be{\bar\eta}

\def\tY{\tilde Y}

\def\lf{16\pi^2}

{\nopagenumbers
\line{\hfil LTH 419}
\line{\hfil hep-ph/9712542}
\vskip .5in
\centerline{\titlefont Renormalisation Invariance and the 
Soft $\beta$-Functions}
\vskip 1in
\centerline{\bf I.~Jack, D.R.T.~Jones and A. Pickering}
\medskip   
\centerline{\it Dept. of Mathematical Sciences,
University of Liverpool, Liverpool L69 3BX, UK}
\vskip .3in
We demonstrate that the soft supersymmetry--breaking 
terms  in a $N=1$ theory can be linked by 
simple renormalisation group  invariant relations which are valid 
to all orders of perturbation theory. In the special case of finite $N=1$ 
theories, the soft terms preserve finiteness to all orders. 
\Date{ December 1997}

Recently there has been remarkable progress in the understanding 
of the soft \sy-breaking $\beta$-functions
~\ref\shif{J.~Hisano and M.~Shifman, \prd 56 (1997) 5475}%
\nref\jjg{I.~Jack and
D.R.T.~Jones, hep-ph/9709364}%
--\ref\avd{L.V.~Avdeev, D.I.~Kazakov and I.N.~Kondrashuk, hep-ph/9709397}.
For a $N=1$ supersymmetric gauge theory with 
superpotential 
\eqn\eqf{W(\Phi)={1\over6}Y^{ijk}\Phi_i\Phi_j\Phi_k+
{1\over2}\mu^{ij}\Phi_i\Phi_j,}
we take the soft breaking Lagrangian $L_{SB}$ as follows: 
\eqn\Aba{\eqalign{
L_{SB}(\Ph,W)&=-\left\{\int d^2\th\eta\left({1\over6}h^{ijk}\Ph_i\Ph_j\Ph_k
+{1\over2}b^{ij}\Ph_i\Ph_j+{1\over2}MW_A{}^{\alpha}W_{A\alpha}\right)
+{\rm h.c.}\right\}\cr
&\quad -\int d^4\th\be\eta\bPh^j(m^2)^i{}_j(e^{2gV})_i{}^k\Ph_k.\cr}}
Here $\eta=\th^2$ is the spurion external field and $M$ is the gaugino mass. 
Use of the spurion
formalism in this context was pioneered by Yamada
\ref\yam{Y.~Yamada, \prd50 (1994) 3537}; in \jjg, \avd\ 
it was shown that $\beta_h$, $\beta_b$ and $\beta_M$ are given by 
the following simple expressions:

\eqna\Ai$$\eqalignno{
\beta_h^{ijk}&=\gamma^i{}_lh^{ljk}+\gamma^j{}_lh^{ilk}
+\gamma^k{}_lh^{ijl}-2\gamma_1^i{}_lY^{ljk}
-2\gamma_1^j{}_lY^{ilk}-2\gamma_1^k{}_lY^{ijl} & \Ai a\cr
\beta_b^{ij}&=\gamma^i{}_lb^{lj}+\gamma^j{}_lb^{il}
-2\gamma_1^i{}_l\mu^{lj}-2\gamma_1^j{}_l\mu^{il} &\Ai b\cr
\beta_M&=2{\cal O}\left({\beta_g\over g}\right) &\Ai c\cr}$$
where 
\eqn\Ajb{
{\cal O}=\left(Mg^2{\partial\over{\partial g^2}}-h^{lmn}{\partial
\over{\partial Y^{lmn}}}\right),}
and
\eqn\Ajxxx{
(\gamma_1)^i{}_j={\cal O}\gamma^i{}_j.}
$\gamma^i{}_j (g, Y, Y^*)$ is the anomalous dimension of the chiral multiplet. 
These relations are valid in DRED (\sic\ dimensional regularisation with 
minimal subtraction).
A straightforward application of the spurion formalism 
leads to the result
\eqn\Ajc{
(\beta_{m^2})^i{}_j=\Delta \gamma^i{}_j,} 

where
\eqn\Ajz{
\Delta = 2{\cal O}{\cal O}^* +2M\bM g^2{\partial
\over{\partial g^2}} +\tY_{lmn}{\partial\over{\partial Y_{lmn}}}
+\tY^{lmn}{\partial\over{\partial Y^{lmn}}},}
$Y_{lmn} = (Y^{lmn})^*$, and 
\eqn\Ajd{
\tY^{ijk}=(m^2)^i{}_lY^{ljk}+(m^2)^j{}_lY^{ilk}+(m^2)^k{}_lY^{ijl}.}
This result, however, is not valid in DRED because the $\epsilon$-scalars 
associated with DRED acquire a mass through 
radiative corrections~\ref\jj{I.~Jack and
D.R.T.~Jones, \plb 333 (1994) 372}. 
Moreover there is no scheme perturbatively related to DRED such that 
Eq.~\Ajc\ is valid. It is, however, possible to 
define a scheme, $\hbox{DRED}'$, closely related to DRED, 
such that $\beta_{m^2}$ is independent of the $\epsilon$-scalar mass
\ref\jjmvy{I.~Jack, D.R.T.~Jones, 
S.P.~Martin, M.T.~Vaughn and Y.~Yamada, \prd50 (1994) R5481}.
Let us hypothesise that in $\hbox{DRED}'$  
the correct result for $\beta_{m^2}$ is 
\eqn\Ajy{
(\beta_{m^2})^i{}_j=\left[ \Delta 
+ X(g, Y, Y^*, h, h^*, m, M)\frakk{\pa}{\pa g}\right]\gamma^i{}_j,} 
where the $X$ term represents in some way the contribution of 
the $\epsilon$-scalar mass renormalisation.
From the explicit calculation of~\jj, \jjmvy\ 
we know that in $\hbox{DRED}'$ the 
leading contribution to $X$ is given by   
\eqn\Ajx{X = -2Sg^3(\lf)^{-1}}
where 
\eqn\Awc{
S\delta_{AB}=(m^2)^k{}_l(R_AR_B)^l{}_k
-MM^* C(G)\delta_{AB}}
(see~\jj\ for our group theory conventions). 
In this paper we will show that the existence of the X-term is 
a necessary consequence of a quite different hypothesis: namely, the 
existence of a set of renormalisation group invariant relations 
expressing $Y$ as a function of $g$;  and $h, b$ and $m^2$ as functions 
of $M$, $\mu$ and $g$. These relations amount to a generalisation to 
the softly-broken case of the coupling constant reduction 
program~\ref\zimm{N.-P.~Chang, \prd10 (1974) 2706\semi
N.-P.~Chang, A.~Das and J.~Perez-Mercader, \prd 22 (1980) 1829\semi
R.~Oehme, K.~Sibold and W.~Zimmermann, \plb 153 (1985) 142\semi
R.~Oehme and W.~Zimmermann,  Comm. Math. Phys. 97 (1985) 569\semi
W.~Zimmermann, {\it ibid} 97 (1985) 211\semi
R.~Oehme,  Prog. Theor. Phys. Suppl. 86 (1986) 215\semi
R.~Oehme, hep-th/9511006\semi
J.~Kubo, M.~Mondrag\'{o}n and G.~Zoupanos, \plb 389 (1996) 523\semi
J.~Kubo, M.~Mondrag\'{o}n and M.~Olechowski, \npb 479 (1996) 25 }. 
Remarkably, implementation of this program 
will enable us to verify Eq.~\Ajx.

In what follows we will specialise for simplicity 
to the case of a single real 
Yukawa coupling $Y$ and a single superfield $\Phi$ transforming 
according to a representation $R$ of the gauge group, 
which we assume admits both  a cubic and a quadratic 
invariant, and we will take all the 
soft terms to be real as well.  
Let us suppose that there exists a RG invariant 
trajectory $Y(g)$. It follows immediately that
\eqn\newa{\beta_Y = 3\ga Y = Y'\beta_g, \quad\hbox{where}\quad 
Y' = \frakk{dY}{dg}.}   
If we seek a perturbative solution to Eq.~\newa\ of the 
form 
\eqn\newb{Y = ag + bg^3 + cg^5 + \cdots}
then we have\ref\jjb{I.~Jack and
D.R.T.~Jones, \plb 349 (1995) 294} 
\eqn\newc{a^2 = 4C(R) + 2Q/3 \quad\hbox{and}\quad b=0}
where $\beta_g = Qg^3(\lf)^{-1}+\cdots$. The special case $Q = a^2 - 4C(R) = 0$ 
corresponds to a one--loop finite theory; it is possible to 
construct $Y(g)$ then so that $\beta_g = \gamma = 0$ to all orders 
\ref\lpsk{C.~Lucchesi, O.~Piguet and K.~Sibold, \plb 201 (1988) 241\semi
C. Lucchesi, hep-th/9510078\semi
A.V.~Ermushev, D.I.~Kazakov and O.V.~Tarasov, \npb 281 (1987) 72\semi
D.I. Kazakov, \mpla 2 (1987) 663\semi 
D.I. Kazakov, \plb 179 (1986) 352}.
The extension of the finite case to include the soft--breaking terms 
has been considered recently by Kazakov~\ref\kaz{ 
D.I.~Kazakov, hep-ph/9709397}; our results generalise his to 
the case of a non-trivial solution to Eq.~\newa.

It was shown in~\jjb\ 
that, given Eq.~\newc, 
the following relations among the soft parameters are RG invariant through
two loops:
\eqna\Aj$$\eqalignno{h &=-MY,&\Aj a\cr
m^2 &= \frakk{1}{3}(1-{1\over{\lf}}{2\over3}g^2Q)M^2,&\Aj b\cr
b &=-{2\over3}M\mu.&\Aj c\cr}$$
We will now proceed to extend these relations to all orders in $g$.
 
For real $Y$, we have 
\eqn\newd{
{\cal O} = \frakk{1}{2}\left(Mg{\partial\over{\partial g}}
-h{\partial\over{\partial Y}}\right),} 
inspection of which suggests immediately the possibility 
\eqn\newe{
h = -MgY'}
since then we have simply
\eqn\newf{
{\cal O} = \frakk{1}{2}Mg\frakk{d}{dg}.}
In particular, for the finite case such that 
$\ga\left(g, Y(g)\right) = 0$ for any $g$, 
then also $\ga_1 = {\cal O}\ga = 0$ and hence $\beta_h =0$. 
Notice that in the 
approximation $Y = ag$ we have $h = -MY$. Thus Eq.~\newe\ provides  the 
generalisation of Eq.~\Aj{a}\ to all orders. 

Just as we 
wanted the relation $Y= Y(g)$ to define a trajectory rather than a 
point in the space of couplings, thus leading to Eq.~\newa, we also require 
Eq.~\newe\ to be RG invariant, which is true if  
\eqn\newg{\beta_h + \left(\beta_M g + M\beta_g\right)Y' 
+Mg\beta_g Y'' = 0.}
It is easy to verify Eq.~\newg, using Eqs.~\Ai{a}\ and \Ai{c}.

The corresponding result for the soft $\phi^2$ mass term $b$ is 
\eqn\newh{b = -\frakk{2}{3}\mu M\frakk{g}{Y}Y'}
and this can also be shown to be RG invariant in a similar way. 
If we assume that there also exists a RG trajectory for $\mu$, 
of the form $\mu = \mu(g)$ then Eq.~\newh\ can be written
\eqn\newha{ b= -Mg\mu',}
which is similar to Eq.~\newe, and also generalises more easily to 
the many-coupling case, which we shall discuss later.

We turn now to the $\beta$-function for the soft $\phi\phi^*$ mass term. 
As we already indicated this presents special problems. We will seek 
a RG invariant relation of the form 
\eqn\newi{m^2 = \frakk{1}{3}M^2 f(g),}
the form of which is motivated 
by Eq.~\Aj{b}. 

For real $Y$ we obtain from Eq.~\Ajb\ and \newe\ that
\eqn\newj{{\cal O}{\cal O^*} = 
\frakk{M^2}{4}\left[g^2\frakk{\pa^2}{\pa g^2}+g\frakk{\pa}{\pa g} 
+2g^2Y'\frakk{\pa^2}{\pa g\pa Y} 
+g^2\left(Y'\right)^2\left(\frakk{\pa^2}{\pa Y^2} 
+ \frakk{1}{Y}\frakk{\pa}{\pa Y}\right)\right],}
and hence (using Eqs.~\Ajy, \newa) that
\eqn\newk{\eqalign{
\beta_{m^2} &= M^2\biggl[ \frak{1}{2}g^2\ga'' + \frak{1}{2}g\ga' 
-\frak{1}{2}\left( g^2 Y'' +gY' 
- g^2\frakk{(Y')^2}{Y}\right)\frakk{\pa\ga}{\pa Y}\cr
& +(g + \Xtilde)\frakk{\pa\ga}{\pa g} + f(g) Y\frakk{\pa\ga}{\pa Y}\biggr],\cr}}
where we have written $X = M^2\Xtilde(g)$ on the RG trajectory.

Now in order that we can obtain $\beta_{m^2}=0$ in the finite case 
it is clear that the partial derivatives with respect to $g$ and $Y$ 
in the above expression will need to fit together into total derivatives 
with respect to $g$. Thus we require
\eqn\newl{
f(g) = \left(\frakk{3}{2}g + \Xtilde\right)\frakk{Y'}{Y} 
+\frakk{1}{2}g^2\left[\frakk{Y''}{Y}- \left(\frakk{Y'}{Y}\right)^2\right]}
whence 
\eqn\newm{\beta_{m^2} = M^2\left[\frakk{1}{2}g^2\ga'' + 
\left(\frakk{3}{2}g + \Xtilde\right)\ga'\right].}

By demanding RG invariance of Eq.~\newi, however, 
we obtain another expression for $\beta_{m^2}$:
\eqn\newj{\beta_{m^2} = M^2\frakk{1}{3}\left[\left(f'-2f/g\right)\beta_g +
2f\beta'_g\right].}

It follows that $\Xtilde$ satisfies the equation
\eqn\newn{g^2\frakk{d}{dg}\left(\frakk{\Xtilde\beta_g}{g^2}\right) = 
\Xtilde'\beta_g - \frakk{2}{g}\Xtilde\beta_g +\Xtilde\beta'_g = 
\frakk{1}{2}\left(3\beta_g-3g\beta'_g + g^2\beta''_g\right)}
whence 
\eqn\newp{X = M^2\Xtilde = 
M^2\left[\frakk{1}{2}g^2 \frakk{\beta'_g}{\beta_g}-\frakk{3}{2}g + 
A\frakk{g^2}{\beta_g}\right]}
where $A$ is an arbitrary constant. From Eq.~\newl\ we then find  
\eqn\news{\eqalign{f(g) &= 
\frakk{1}{2}g^2\left[\frakk{Y''}{Y}- \left(\frakk{Y'}{Y}\right)^2
+ \frakk{\beta'_gY'}{\beta_g Y}\right]+ A\frakk{g^2Y'}{\beta_g Y}\cr
&= \frakk{3g^2}{2\beta_g}\ga' + A\frakk{g^2Y'}{\beta_g Y}.\cr}}

We can now compare our result for $X$ with the existing 
perturbative calculation~\jj\ of $\beta_{m^2}$.  Now we know that at the 
one loop level, $\beta_{m^2}$ satisfies Eq.~\Ajy\ with $X=0$; we expect 
the leading contribution to $X$ to be $O(g^3)$. It follows that we must take 
$A=0$ above. Using Eq.~\newc\ and the two--loop 
result for $\beta_g$~\ref\two{A.J. Parkes and P. West, \plb138 (1984) 99;
             \npb256 (1985) 340\semi
             P. West, \plb137 (1984) 371\semi
          D.R.T. Jones and L. Mezincescu, \plb136 (1984) 242;
{\it ibid} 138 (1984) 293.}\ we obtain
\eqn\newq{\beta_g = Qg^3(\lf)^{-1} - \frakk{2}{3}Q^2 g^5(\lf)^{-2}+\cdots}
and hence 
\eqn\newr{X = -\frakk{2}{3}QM^2g^3(\lf)^{-1}+ \cdots}
while from Eq.~\news\ we obtain
\eqn\newt{f(g) = 1-{1\over{\lf}}{2\over3}g^2Q+ \cdots.}
Observe that this result for $f$ is consistent with Eq.~\Aj{b}; moreover, 
it is easy to show that our result for $X$ is consistent with 
Eq.~\Ajx. 

Let us now consider the 
special case of a finite theory, already considered in \kaz. 
 If we define $h$ by Eq.~\newe, then it follows 
immediately from Eqs.~\Ai{}, \Ajxxx\ 
and \newf\ that if $\beta_g = \ga =0$ then 
$\beta_h = \beta_M = 0$ to all orders. We also have $\beta_b = 0$; 
notice that this result in fact is true even if we do not impose 
Eq.~\newh; which is why Eq.~\newh\ does not appear in \kaz. 
Finally, from Eq.~\newm, we see that in the finite case 
we have $\beta_{m^2} = 0$ to all orders as long as 
$X$ is well-defined in that case. This is not quite a trivial requirement, 
as can be seen from Eq.~\newp. But in the finite case we have 
\eqn\newu{\beta_g(g, Y(g)) = Q g^3F(g)(\lf)^{-1}}
where $F(g) = 1 - \frakk{2}{3}Qg^2(\lf)^{-1} + \cdots,$ whence it follows that 
$X$ is finite when $\beta_g = 0$. There is, however, no reason to think 
that $X$ is zero to all orders, in the finite case, although the leading 
contribution vanishes, as can be seen from Eq.~\newr. 

Thus we have shown that \sic\ theories including soft terms
admit RG invariant trajectories for both the Yukawa couplings 
and the  soft terms. As a consequence the formula for $\beta_{m^2}$  
requires a term not predicted by a naive application of the spurion 
formalism. We have determined the associated RG function to all orders 
on the aforesaid trajectory:

\eqn\conca{
X = M^2\left[\frakk{1}{2}g^2 \frakk{\beta'_g}{\beta_g}-\frakk{3}{2}g\right].} 
Our results for the soft terms are:
\eqna\concb$$\eqalignno{h  &= -M g\frakk{dY}{dg}&\concb a\cr
b &= -Mg\frakk{d\mu}{dg} &\concb b \cr
m^2 &= \frakk{g^2}{2\beta_g}M^2\frakk{d\ga}{dg}.&\concb c\cr}$$

Let us now discuss the case of a general superpotential, Eq.~\eqf.
In Eq.~\conca\ we have merely to replace $M^2$ by $MM^*$, while in place
of Eq.~\concb{}\ we have:
\eqna\concc$$\eqalignno{h^{ijk}  &= -M g\frakk{dY^{ijk}}{dg} &\concc a\cr
b^{ij} &= -Mg\frakk{d\mu^{ij}}{dg} &\concc b\cr
(m^2)^i{}_j &= \frakk{g^2}{2\beta_g}MM^*\frakk{d\ga^i{}_j}{dg}. &\concc c\cr
}$$
In deriving Eq.~\concc{}\  we have used the generalisation of 
Eq.~\newa, 
\eqn\concca{\beta_Y^{ijk} = Y^{l(ij}\ga^{k)}{}_l = \frakk{dY^{ijk}}{dg}\beta_g.}
We found it necessary
to assume that
\eqn\concd{
Y^{ijk}\frakk{\pa\ga^l{}_m}{\pa Y^{ijk}} 
= Y^{*ijk}\frakk{\pa\ga^l{}_m}{\pa Y^{*ijk}},
 \quad(\hbox{no sum on $i,j,k$})}
and a similar equation for $\beta_g$, 
and also that $\ga$ is diagonal.  Using these  assumptions and 
Eq.~\concca, it is easy to show that  
\eqn\concd{
Y^{\prime ijk}\frakk{\pa\ga^l{}_m}{\pa Y^{ijk}} 
= Y^{*\prime ijk}\frakk{\pa\ga^l{}_m}{\pa Y^{*ijk}},
 \quad(\hbox{no sum on $i,j,k$}),}
and once again a similar equation for $\beta_g$,
which is necessary, for example, 
to establish Eq.~\newf\ in the general case.

The fact that in the general case the soft terms preserve finiteness
 in finite \sic\ theories also follows from Eq.~\concc{}. 
Here we are broadly in agreement with~\kaz, 
but we believe our analysis places the results on a firmer 
footing, being associated with a specific and well-defined subtraction 
procedure. 

An interesting question left unanswered is the form of
$X$  away from the RG invariant trajectory $Y = Y(g)$. We hope to return
to this, and the phenomenological  consequences of our general 
result Eq.~\concc{}, elsewhere.  

\bigskip\centerline{{\bf Acknowledgements}}\nobreak

AP was supported by a PPARC Research Grant. 

\listrefs
\bye